\documentclass{jpp}
\usepackage{graphicx}
\usepackage{physics}
\usepackage[utf8]{inputenc}
\usepackage[T1]{fontenc}
\usepackage{amsmath}
\usepackage{float}
\usepackage{hyperref}
\usepackage[dvipsnames]{xcolor}
\usepackage{siunitx}

\shorttitle{Drift of ablated material after pellet injection in a tokamak}
\shortauthor{O. Vallhagen and others}

\title{Drift of ablated material after pellet injection in a tokamak}
\author{O.~Vallhagen\aff{1}, I.~Pusztai\aff{1},
P.~Helander\aff{2},
S.~L.~Newton\aff{3},
 \and   T.~F\"{u}l\"{o}p\aff{1}}
\affiliation{\aff{1}Department of Physics, Chalmers University of Technology, G\"{o}teborg, SE-41296, Sweden
\aff{2} Max-Planck Institute for Plasma Physics, Greifswald, Germany
\aff{3}  Culham Centre for Fusion Energy, Abingdon, Oxon OX14 3DB, United
Kingdom
}
\begin{document}

\maketitle

\begin{abstract}
Pellet injection is used for fuelling and controlling discharges in tokamaks, and it is foreseen in ITER. During pellet injection, a movement of the ablated material towards the low-field side (or outward major radius direction) occurs because of the inhomogeneity of the magnetic field.  Due
to the complexity of the theoretical models, computer codes developed to simulate the
cross-field drift are computationally expensive. Here, we present a one-dimensional semi-analytical model for the radial displacement of ablated material after pellet injection, taking into account both the Alfv\'en and ohmic currents which short-circuit the charge separation creating the drift. The model is suitable for rapid calculation of the radial drift displacement, and can be useful for e.g.~modelling of disruption mitigation via pellet injection. 
\end{abstract}

\section{Introduction}
Pellet injection  is an effective tool for modifying the density profile in fusion devices, and can be used for both fuelling and plasma control \citep{Pegourie2007}. It has also been employed successfully to mitigate transient events in tokamaks, e.g.~edge localized modes \citep{Lang_2015} and disruptions \citep{Reux2021}. The use of pellets to control such events is also planned for ITER \citep{Baylor_2009,Hollmann2015,lehnen_2018}. 

In order to assess the performance of pellet injection schemes for future tokamaks, such as ITER, it is important that accurate estimates of the modified density profile created by the pellets are included in the modelling tools used to simulate such events. This can only be achieved through an understanding of the underlying physics of the mass deposition after pellet injection.

When a pellet is injected into a hot, magnetically confined plasma, it travels through the plasma in solid form while the outer layers are continuously ablated by the energy flux from the hot background plasma, resulting in material being deposited along the pellet trajectory. The cloud of ablated material initially has a cold dense structure -- a plasmoid -- which drifts towards the low-field side of the torus. This is caused by the charge separation that takes place due to electron and ion drifts in the inhomogeneous magnetic field, leading to the build-up of a vertical electric field, and the resulting $\mathbf{E}\times \mathbf{B}$-drift moves the ablated material across magnetic field lines \citep{Parks2000,Rozhansky2004, Pegourie2006}. 

The strength of the electric field, and hence the drift velocity, is determined by the mechanisms which can short-circuit the charge separation inside the plasmoid. The dominant ones are the emission of Alfvén waves from the two ends of the plasmoid \citep{Parks2000} and the flow of ohmic current parallel to the field lines \citep{Pegourie2006}. Mathematically, the evolution of the pellet cloud is governed by a vorticity equation similar to that used to describe so-called blob transport in the plasma scrape-off layer \citep{Krasheninnikov_2008}.

There is a wealth of experimental evidence for radial cross-field drift following pellet injection in current tokamaks and stellarators, e.g.~in DIII-D \citep{Baylor_2007}, ASDEX Upgrade \citep{Lang1997, Muller1999}, FTU \citep{Terranova_2007}, MAST \citep{Garzotti_2010} and W7-X \citep{Baldzuhn_2019}. However, these studies consider small fuelling pellets, and there is much less experimental data on radial drifts from strongly perturbing pellets used for disruption mitigation, although there are recent indications that radial drifts may be important also in such cases at DIII-D and JET \citep{Lvovskiy_APS2022, Kong_EPS2022}. Due to the complexity of the theoretical models, computer codes developed to simulate the cross-field drift are computationally expensive \citep{Strauss1998,Strauss_2000,Aiba_2004,Ishizaki_2011, Samulyak_2021}. Therefore, simplified scaling laws, based on current experimental observations, are often used \citep{Baylor_2007,koechl_2018}. Such expressions are of limited use for modelling ITER plasmas, which will have much higher temperatures and magnetic fields. In many cases, e.g.~in the currently used disruption mitigation models, the radial drift of the pellet cloud is neglected altogether, for simplicity \citep{Vallhagen2022}. This is particularly problematic in the case of pure hydrogen pellets \citep{Matsuyama2022}, as their clouds can reach significant over-pressure due to negligible radiative energy losses, thus their drifts can be large and therefore affect the pellet penetration and material deposition substantially.

The purpose of this paper is to develop a semi-analytical model for the cross-field drift motion of the ionized plasmoid, taking into account both the Alfv\'en and ohmic currents. Our aim is to extract the key physical mechanisms described by the codes mentioned above and condense the result into a computationally efficient model. We consider current conservation directly, rather than formulating a vorticity equation for the system, generalising the description of the parallel connection of the ohmic current, and clarifying elements present in the existing literature. Factors such as the assumed shape of the plasmoid and  our neglect of its structure along the magnetic field will quantitatively affect the plasmoid dynamics, but will not affect the qualitative nature of the results presented here.

\section{Physical model}
The motion of the plasmoid arises because of an $\mathbf{E}\times \mathbf{B}$-drift in the direction of the major radius; the electric field builds up due to the current from the magnetic (curvature + $\nabla B$) drift of the particles, while the time-variation of this electric field gives rise to a partially cancelling polarization drift current. The total radial shift is determined by the drift velocity reached and its duration, which is approximately the time it takes for the cloud to expand one connection length along the field lines ($t\sim \pi R_\mathrm{m} q/c_s$, where $R_\mathrm{m}$ is the major radius, $c_s$ is the sound speed and $q$ is the safety factor or inverse of the rotational transform of the magnetic field). At this time, magnetic drift currents in the outboard and inboard portions of the cloud cancel out (analogously to a tokamak equilibrium).
    
In order to mathematically describe the pellet dynamics, we formulate the current-conservation equation for the system, describing the balance between the divergent parts of the currents necessary to maintain quasineutrality.
Working within a single-fluid formalism, we introduce the mass density $\rho$, the mass flow velocity $\mathbf{v}$, which appears in the total time derivative $d_t=\partial_t+\mathbf{v}\cdot\nabla$, the total pressure including the electron and ion pressure components $p=p_e+p_i$, as well as the current density and the magnetic field vectors, $\mathbf{j}$ and $\textbf{B}$. In addition, $\mathbf{B}=\mathbf{b}B$ with the unit vector $\mathbf{b}$, the curvature vector of the field lines is $\boldsymbol{\kappa}=\mathbf{b}\cdot \nabla \mathbf{b}$, and $\mu_0$ denotes the vacuum permeability. 

The pellet cloud has higher pressure than the surrounding plasma since it is continuously heated by hot electrons from the latter \citep{Parks_Thurnbull}. A current perpendicular to the magnetic field lines arises in response to this excess pressure, but we note that the dynamics involved in the drift of the plasmoid is slower than the timescale of compressional Alfv\'{e}n waves, so that the largest terms in the magnetohydrodynamic (MHD) force balance equation, 
\begin{equation}
\rho \frac{d {\mathbf{v}}}{dt} =\mathbf{j}\times \mathbf{B}-\nabla p,
\label{eqmotion}
\end{equation}
describe an approximately static force balance between the plasma pressure and the magnetic field.
The total current takes the form
\begin{equation}
\mathbf{j} = j_\|\mathbf{b} + \frac{\mathbf{B}\times \nabla p}{B^2} + \frac{\rho}{B}\mathbf{b}\times\frac{d {\mathbf{v}}}{dt},
\end{equation}
and the divergence of the diamagnetic current, the second term on the right, describes the charge accumulation due to the magnetic drifts (driven by field line curvature and field strength inhomogeneity).
This is approximately given by
\begin{equation}
\nabla \cdot \left(\frac{\mathbf{B}\times \nabla p}{B^2}\right) = \nabla \cdot \left[ p \nabla \times \left(\frac{\mathbf{B}}{B^2}\right)\right] \approx \nabla \cdot \left( 2p\frac{\mathbf{b} \times \nabla B}{B^2}\right) \equiv \nabla \cdot \mathbf{j}_{\nabla B}.
\end{equation}

Using $
(\mathbf{b} \times \nabla B)/B = \mathbf{b} \times\boldsymbol{\kappa}$ we can write the expression for current conservation in the form
\begin{equation}
0 = \nabla \cdot \mathbf{j} \approx \nabla \cdot \left[j_\|\mathbf{b} + \frac{\mathbf{b}}{B} \times\left( 2p \boldsymbol{\kappa} +\rho \frac{d {\mathbf{v}}}{dt} \right)\right].
\label{divergencefree}
\end{equation}
The time-dependent term in~(\ref{divergencefree}) is the current due to the polarization drift,
$$\mathbf{j}_{\dot{\mathbf{E} }}=\rho \frac{\mathbf{b}}{B}\times \frac{d \mathbf{v}}{dt}.$$
The resistive-MHD Ohm's law $\mathbf{E}+(\mathbf{v}\times \mathbf{B})=\eta \mathbf{j}$ implies that in the limit of modest resistivity $\eta$, the perpendicular mass flow $\mathbf{v}_\perp$ is dominated by $\mathbf{E}\times \mathbf{B}$ motion. 
For the low-frequency process of interest inside the pellet cloud, the electric field is electrostatic $\mathbf{E}=-\nabla \phi$, with the electrostatic potential $\phi$, and thus we write the cross-field velocity as $\mathbf{v}_\perp\approx (\mathbf{b}\times \nabla \phi)/B$. 

The parallel current ($j_\|$) must adjust to make the total current divergence free; that is $$\nabla\cdot \mathbf{j}=\nabla \cdot[j_\| \mathbf{b}+\mathbf{j}_{\nabla B}+\mathbf{j}_{\dot{\mathbf{E} }}]=0.$$
Eliminating the contribution which describes the balance of the diamagnetic and parallel currents in the background equilibrium plasma, we are left with the perturbation of the current continuity equation driven by the excess pressure of the plasmoid. 

In the very early phase of plasmoid acceleration, $\mathbf{j}_{\nabla B}$ is approximately balanced by $\mathbf{j}_{\dot{\mathbf{E} }}$. At such short times, the length of the pellet cloud is much shorter than the distance around the torus, $t\ll R_m q/c_s$, and the plasmoid is thus poloidally and toroidally localised. If the aspect ratio of the torus is large, the curvature vector of the magnetic field is approximately $\boldsymbol{\kappa}=-\hat{R}/R_m$, where $R_m$ is the major radius and
$\hat{R}$ the unit vector in the direction of increasing major radius. For convenience we introduce the unit vector $\hat{Y}=\mathbf{b}\times \hat{R}$, so the direction of $\mathbf{j}_{\nabla B}$ is $-\hat{Y}$, which is nearly vertical.
As the electric field rises in this early stage, $\mathbf{j}_{\dot{\mathbf{E} }}$ evolves to point in the $\hat{Y}$ direction everywhere in the cloud. Later the $j_\|$ term starts to dominate over $\mathbf{j}_{\dot{\mathbf{E} }}$ in balancing $\mathbf{j}_{\nabla B}$, setting the quasi-steady speed of the plasmoid. 

\begin{figure}
    \centering
    \includegraphics[width=0.99\textwidth]{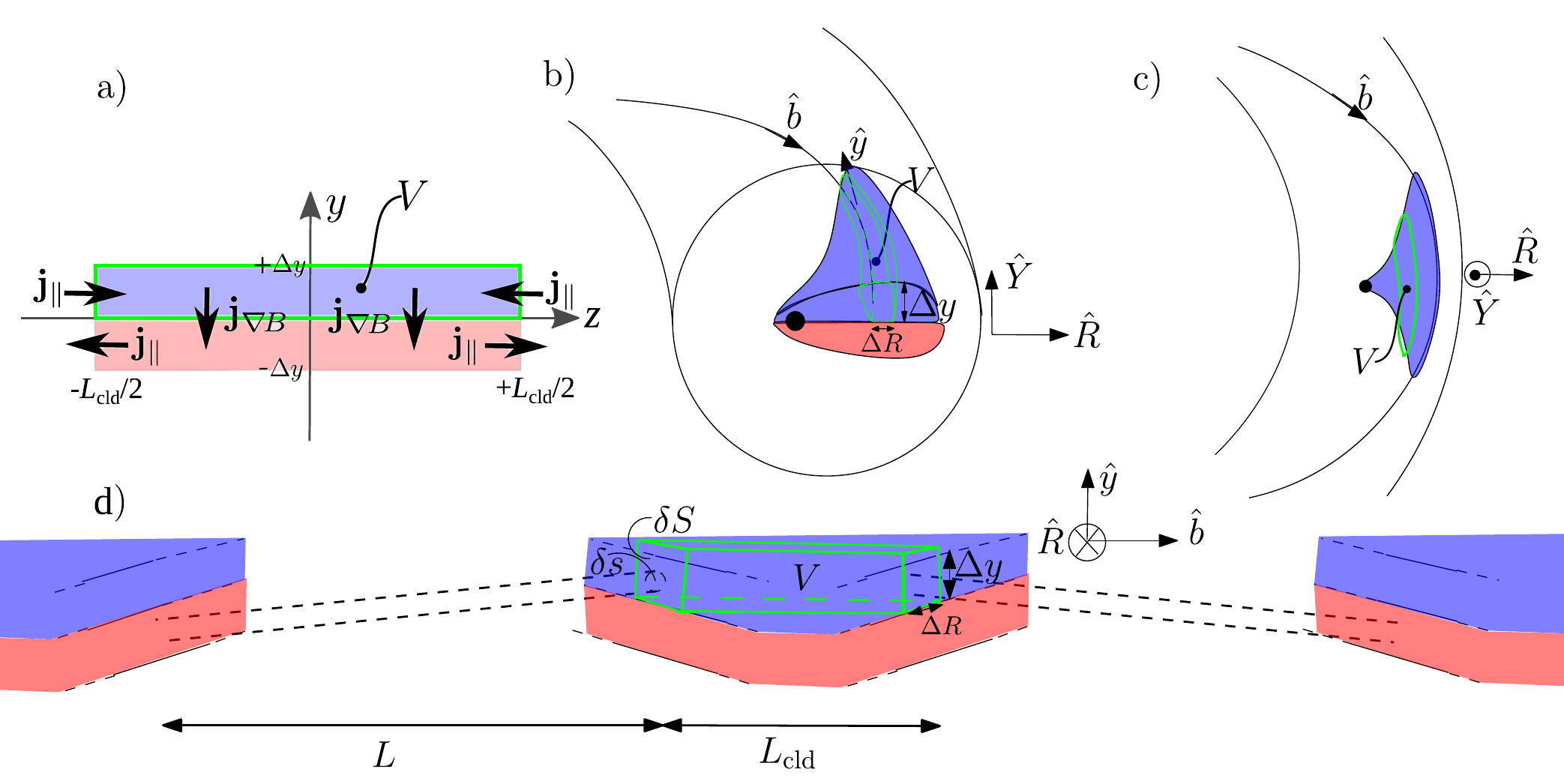}
    \caption{Schematic views of the ablation cloud and the field lines connecting the various parts of it from different perspectives; the green lines indicate the boundaries of the integration volume $V$: a) parallel currents and magnetic drift currents indicated in the $y-z$ plane, b) from the side looking in the toroidal direction, c) from the top and d) with unwrapped field lines (black dashed), connecting different parts of the cloud after a distance $L$.   The cloud expands at the speed of sound $c_s$ in both directions, so that $L_\mathrm{cld}=2 c_s t$. We assume the pellet ablation cloud to be symmetric in $z$ (and $y$) with respect to the $y$ ($z$) axis in figure a). The pellet is indicated in figures b) and c) by the black dot, from which the cloud diverges.}
    \label{fig:geom}
\end{figure}

We may integrate (\ref{divergencefree}) over some convenient volume $V$ with boundary $\partial V$, and apply the divergence theorem to obtain
\begin{align}
        0 = &\int _{\partial V} \left[ \left(\frac{\rho}{B} \frac{d\mathbf{v}}{dt}-\frac{2p}{BR_m}\hat{Y}\right)+  j_{\|}\mathbf{b}\right]\cdot \hat{n}dS,
    \label{integral}
\end{align}
where $\hat{n}$ is a unit vector pointing outwards from $V$. 
 We align the integration volume $V$ with the cloud by choosing it to be a magnetic flux tube extending along the length of the cloud. Since the magnetic field lines are curved, the end faces of the flux tube, which we denote by $\delta S$, are not quite parallel. We choose the flux tube to have rectangular cross section with the lower boundary running through the middle of the cloud, separating the upper, blue, and lower, red, parts of the cloud shown in Fig.~\ref{fig:geom}, where the integration volume $V$ is sketched. The length of the cloud along the field line is $L_{\rm cld}$, and the upper boundary of the domain is located just above the cloud. 
 
 For simplicity, we assume that the pellet is injected in the horizontal midplane and therefore (by symmetry) is always located in the middle of the cloud in the direction along the magnetic field and in the vertical direction. The surface normal $\hat{y}$ of the lower surface of $V$ coincides with $\hat{Y}$ in the poloidal plane that contains the pellet, and rotates in the poloidal plane as one follows the field line along the flux tube $V$. The relation between $\hat{y}$, $\hat{Y}$ and $\hat{R}$ is 
\begin{equation}
    \hat{y} = \cos{\theta} \,\hat{Y} + \sin{\theta}\, \hat{R},
\end{equation}
where $\theta \approx \varphi/q \approx z/qR_\mathrm{m}$ is the poloidal angle, $\varphi$ is the toroidal angle and $z$ is the coordinate along the magnetic field lines; we take $z=0$ in the poloidal plane of the pellet. 
 The dimensions of the integration volume in the $\hat{R}$ and $\hat{y}$ directions are $\Delta R$ and $\Delta y$, respectively.
 
 The contribution from the first term, $\mathbf{j}_{\dot{\mathbf{E} }}$, to \eqref{integral} thus becomes
 \begin{equation}
     I_{\dot{\mathbf{E} }} = \int _{-L_\mathrm{cld}/2}^{L_\mathrm{cld}/2} \int _0^{\Delta R} \frac{\rho}{B^2} \frac{dE_y}{dt}\hat{y}\cdot \hat{y}dRdz = \frac{\bar{n}\langle m_i \rangle\Delta R}{(1+\langle Z \rangle)B^2} \frac{dE_y}{dt},
 \end{equation}
where we have noted that the field-line-integrated mass density is $\bar{n}\langle m_i\rangle/(1+\langle Z \rangle)$ (neglecting the mass of the electrons), $\bar{n} = \sum _{i} \bar{n}_i + \bar{n}_e = \sum _i \bar{n}_i(1+\langle Z \rangle)$ is the field-line integrated total density of all species (including electrons) inside the cloud (with $\bar{n}_i$ and $n_e$ denoting the field line integrated density of ion species $i$ and electrons, respectively), $\langle m_i \rangle$ is the average ion mass inside the cloud and $\langle Z \rangle$ is the average ion charge inside the cloud.

Considering the second term, $\mathbf{j}_{\nabla B}$, we assume that the pressure is constant along the field lines inside the cloud, with equal electron and ion temperatures, denoted by $T$.
The contribution from the second term of \eqref{integral} then becomes 
\begin{align}
        I_{\nabla B} &= \int _{-L_\mathrm{cld}/2}^{L_\mathrm{cld}/2} \int _0^{\Delta R} -\frac{2(p-p_\mathrm{bg})}{BR_\mathrm{m}} \hat{Y}\cdot\hat{y}dRdz = \int _{-L_\mathrm{cld}/2}^{L_\mathrm{cld}/2} -\frac{2(p-p_\mathrm{bg})\Delta R}{BR_\mathrm{m}}\cos{\left(\frac{z}{qR_\mathrm{m}}\right)}dz \nonumber\\
        &= -\frac{4(p-p_\mathrm{bg})\Delta Rq}{B}\sin{\left(\frac{L_\mathrm{cld}}{2qR_\mathrm{m}}\right)} = -\frac{4(\bar{n}T-L_\mathrm{cld}n_\mathrm{bg}T_\mathrm{bg})\Delta Rq}{BL_\mathrm{cld}}\sin{\left(\frac{L_\mathrm{cld}}{2qR_\mathrm{m}}\right)},
        \end{align}
The background pressure $p_\mathrm{bg}$ enters via the contribution from the upper surface of the integration volume.
We see, as noted in the introduction, that the assumptions simplifying the parallel structure of the cloud will quantitatively affect the final results, but accounting for parallel structure will not affect the essential qualitative description of the plasmoid motion.
 
The key to calculating how the parallel current contributes to the drift motion is to find the relation between the parallel current $j_{\|}$, which flows through the background plasma (beyond the ends of the cloud), and the electric field responsible for $\mathbf{E}\times \mathbf{B}$ motion, which are related via the electrostatic potential $\phi$ along the plasmoid length. 
As the pellet flies through the plasma, it undergoes continuous ablation and thus generates a sequence of ablation clouds residing on different field lines. Each of these clouds expands along the magnetic field whilst drifting across it. It is important to note that the cloud drift velocity exceeds the speed of the pellet. We can thus regard the pellet as stationary, which simplifies our discussion. 

With these facts in mind, we now study the evolution of the electrostatic potential along each field line. We fix our attention on one particular field line and denote by $\tau$ the time that has elapsed since pellet material first arrived there. This time is in general different from the time $t$ that has passed since this material was originally ablated from the pellet. (Alternatively, in the limit of very high electrical conductivity, it is possible to regard the field lines as ``frozen into'' the pellet cloud, in which case it is better to consider a field line moving with the pellet cloud. In this case $t=\tau$.)

It is convenient to introduce $L$, the distance along a field line, outside the cloud, which connects the two ends of the cloud; note that $L$ depends on the coordinates identifying a field line and may be different for different field lines in our integration volume $V$. In our large-aspect-ratio approximation, the value of $L = 2 \pi R_\mathrm{m} N$ is equal to the circumference of the torus, $2\pi R_\mathrm{m}$, times the number of turns, $N$, after which the field line connects the two end caps of $V$. This number will in general vary over the cross section of the flux tube. 

 \begin{figure}
    \centering
    \includegraphics[width=0.85\textwidth]{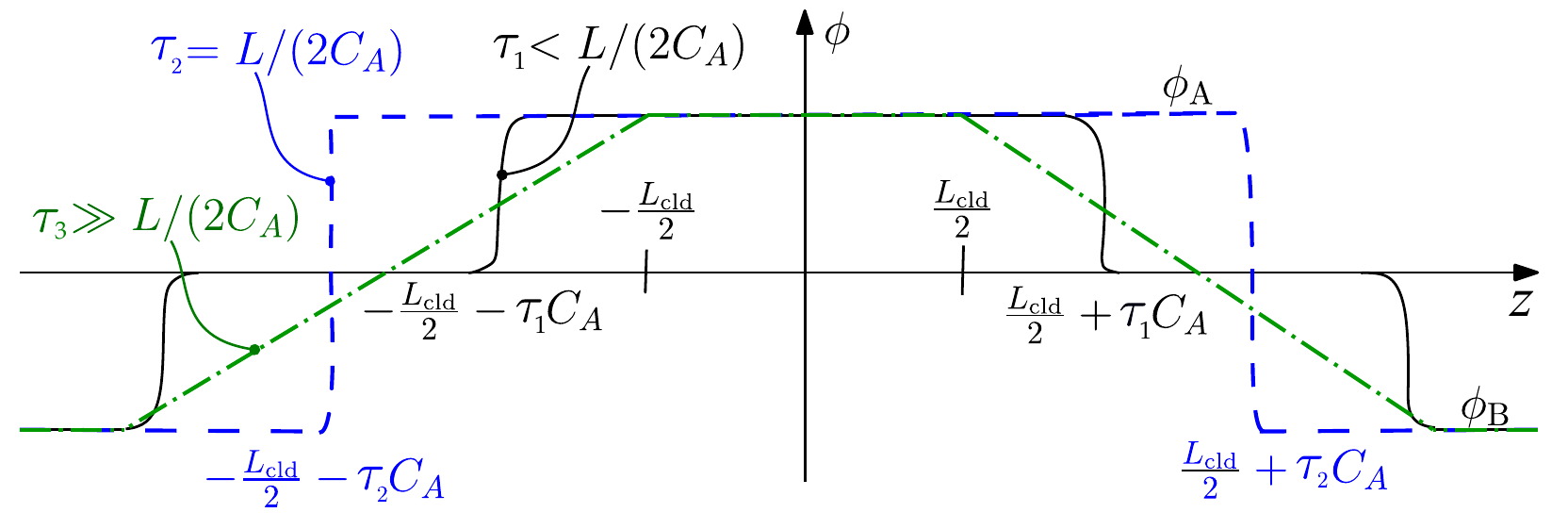}
    \caption{Sketch of the electrostatic potential $\phi(z)$ along a field line connecting the two ends of the cloud, at different values of $y$, characterised by potentials $\phi_A$ and $\phi_B$. We show three representative times: At $\tau_1<L/(2C_A)$ potential perturbations propagating out from the ends of the cloud at the Alfv\'en speed have not yet met along the field line (solid black line). The perturbations meet at $\tau_2=L/(2C_A)$ (dashed blue). After a long time (compared to Alfv\'en time scales), $\tau_3\gg L/(2C_A)$, the potential has reached a quasi-steady state where an ohmic current flows between the connected ends of the cloud (dash-dotted green). Note that the cloud length $L_{\rm cld}$ is exaggerated in the figure; in reality it is much shorter than the distance along the field line between the connected ends of the cloud.}
    \label{fig:Drift_potential}
\end{figure}

The evolution of the electrostatic potential along a field line connecting the oppositely charged parts of the cloud after a length $L$ is illustrated in figure \ref{fig:Drift_potential}. The physical picture of the evolution of this potential is the following: the interface between the end of the plasmoid and the background plasma represents an evolving perturbation, expanding along the field lines at the local sound speed $c_s$ of the pellet material inside the plasmoid. The potential difference between the cloud and the background plasma, along with the plasmoid drift, excite shear Alfv\'en waves, which are emitted from these interfaces and propagate away from the plasmoid, along field lines through the background plasma, at the local Alfv\'en speed, $C_A$.
For $\tau\ll L/(2 C_A)$, the potential perturbations associated with the Alfv\'en waves will not have reached each other yet. Thus, the current carried away from the ends of the cloud is determined by the polarisation current resulting from the time-varying potential at the wave fronts, giving rise to the Alfv\'en current \citep{Scholer1970}.
When $\tau=L/(2 C_A)$, the waves emerging from the opposite sides of the cloud meet and interfere with each other. Eventually, a steady-state, without propagating waves, is reached when $\tau\gg L/(2 C_A)$. At this stage, the parallel current is instead determined by Ohm's law.

  Thus, the dominant contribution to the $j_{\|}$ current, in the initial phase, is associated with the Alfv\'{e}n wave propagating from the ends of the drifting cloud \citep{Parks2000}. It is proportional to the electric field inside the cloud, as outlined below, and can be described by the so-called Alfv\'en conductivity, $\Sigma _A = 1/ R_A = 1/(\mu _0 C_A)$. 
    In the later stages, the ohmic current along the field lines connecting the oppositely charged parts of the cloud \citep{Pegourie2006} becomes dominant. There is also a contribution to the current caused by the drift resulting from the cloud viscosity, which has been shown by \cite{Rozhansky2004} to be less significant and will be neglected here. 
 
   \subsection{Parallel current}

When calculating the contribution of $j_{\|}$ to the integral (\ref{integral}), only the end caps (area $\delta S$) of this flux tube will contribute, as otherwise $\mathbf{b}\cdot \hat{n} = 0$. 
 Consider first the contribution from a smaller flux tube, whose end caps have area $\partial s_<$, that only contains field lines for which $C_A t \ll L/2$, that is, for which Alfv\'en waves propagating from the ends of the cloud have not had time to meet. 
  As the parallel electric field $E_{\|}$ is small in the established hot background plasma outside the cloud, except at the wave front, we can express $E_{\|}$ in Fourier space as $$E_{\|} = -ik_{\|} \phi + i\omega A_{\|} \approx 0,$$ and relate the electrostatic and vector potential via the Alfv\'en speed $$A_{\|} = \phi/(\omega/k_{\|}) = \phi/C_A.$$ Using Amp\`ere's law we can relate $j_{\|}$ to $A_{\|}$ and thence to $\phi$ as
\begin{equation}
    j_{\|} = -\frac{\nabla _\perp^2 A_{\|}}{\mu _0} = -\frac{\nabla _\perp^2 \phi}{\mu _0C_A}.
\end{equation}
Assuming that the whole cloud moves at the same radial velocity, the electric field $E_y = -\partial \phi/\partial y$ must be constant inside the cloud, i.e.
\begin{equation}\nabla _\perp^2 \phi 
= -E_y \left[ \delta(y-\Delta y) - \delta(y+\Delta y) \right], 
\end{equation} 
where $\delta$ denotes the Dirac delta function. If we set $\partial s_<$ to the part of $\partial S$ for which $C_A t <L/2$, the contribution from the Alfv\'en part of the parallel current becomes
\begin{align}
    I_{\|,A} &= 2\int _{\partial s_<} \frac{E_y\delta(y-\Delta y)}{\mu _0C_A}dydR \\ \nonumber
    &= 2\int _0^{\Delta R}\int_0^{\Delta y} \Theta(\partial s_<;y,R)\frac{E_y\delta(y-\Delta y)}{\mu _0C_A}dydR \\ \nonumber
    &= 2 P_A \Delta R \frac{E_y}{\mu _0C_A} = 2 P_A \Delta R \frac{E_y}{R_A} ,
\end{align}
where the function $\Theta(\partial s;y,R)$ is $1$ for the $y$ and $R$ values corresponding to field lines crossing the surface $\partial s$ where $C_A t <L/2$ is satisfied, and zero otherwise; and $P_A$ is the fraction $\partial s_< /\partial S$.   

Now consider the field lines crossing the area $\partial s_>$, i.e., the field lines for which $C_A t\gg L/2$. On these field lines, the Alfv\'en waves emanating from either side of the cloud have already met and decayed, and there is no longer any polarisation current. Only the ohmic current $j_{\|}$ remains and, being divergence free, it must be constant along the field in the large-aspect-ratio limit. This current is related to the parallel electric field by Ohm's law,
$$j_{\|} = \sigma_{\|} E_{\|} = -\sigma_{\|}\nabla _{\|}\phi.$$
As $j_{\|}$ is constant along the field lines, so is $\nabla \phi$, which means that
\begin{equation}
    j_{\|} = \sigma_{\|} E_{\|} = -\sigma_{\|} \frac{\phi _B -\phi}{L} = -\sigma_{\|} \frac{E_y (y_B-y)}{L},
\end{equation}
where $y$ denotes the vertical coordinate at which the field line emanates from one end of the cloud and $y_B$ that where it hits the other end. The electric field $E_y$ has been assumed to be constant along the field line.

Let us now denote by $\partial s_i$ the subset of $\delta s_>$ containing only field lines connecting to the opposite side of the cloud after a distance $L = 2\pi R_\mathrm{m} i$, i.e. connecting after exactly $i$ toroidal turns. If $i \gg 1$, the connection is essentially random, so that the values of $y$ and $y_B$ are uncorrelated and $\int y_Bdy = 0$. The total ohmic current flowing along field lines in $\delta s_i$ thus becomes
\begin{align}
    I_{\|,\mathrm{ohm}}^{(i)}(\tau\gg L/(2C_A))& = -2\int _{\partial s_i} \sigma _{\|} \frac{E_y(y_B - y)}{L}dydR\nonumber\\ &= -2P_i \int _0^{\Delta y} \int _0^{\Delta R}  \Theta(\partial s_>;y,R)  \sigma _{\|} \frac{E_y(y_B - y)}{2\pi R_\mathrm{m} i}dydR \nonumber\\&=  P_i\sigma _{\|} \frac{E_y\Delta y^2\Delta R}{2\pi R_\mathrm{m} i},\label{iohmic}
\end{align}
where $P_i = \delta s_i / \partial S$ is the fraction of the cloud connecting to the opposite side after $i$ toroidal turns. 
This result is similar to the corresponding expression, Eq.~(2), in \citep{Commaux_2010}, up to an order unity factor accounting for the finite electron collision time. 
The total ohmic current is obtained by summing over all values of $i$.

For $\tau\gtrsim L/(2C_A)$, the current will make a transition from 0 to $I_{\|,\mathrm{ohm}}^{(i)}(\tau\gg L/(2C_A))$ \footnote{This only applies to field lines in the interior of $\partial S$; at the boundary of $\partial S$ the initial current is $I_{\|,A}$. However, as the ohmic current is proportional to the cross section area, the boundary of $\partial S$ gives a negligible contribution to the ohmic current.}over a time scale  similar to $L/C_A$, so that we can write
\begin{equation}
    I_{\|, \mathrm{ohm}} = \sum _{i=1}^\infty f\left(\frac{\tau}{L/(2C_A)}\right)I_{\|,\mathrm{ohm}}^{(i)}(\tau\gg L/(2C_A)),
\end{equation}
where $f(0)=0$ and $f\rightarrow 1$ for large arguments. The detailed form of $f$ is determined by the interaction of the Alfv\'en waves propagating from opposite sides of the cloud, which is outside the scope of the present work. Here we instead make the approximation that $f = \theta\left(\frac{\tau}{L/(2C_A)}-1\right)$,  where $\theta$ is the Heaviside step-function.
This is also used in \cite{Pegourie2006}.
Note that this assumption on $f$ underestimates the time until the onset of the ohmic current, thus overestimating the importance of the ohmic current contribution. With this assumption for $f$, we can write the total ohmic current as
\begin{equation}
    I_{\|, \mathrm{ohm}} = \sum _{i=1}^N I_{\|,\mathrm{ohm}}^{(i)}(\tau> L/(2C_A)),\label{eq:iohmic_simple}
\end{equation}
where $N = \lfloor 2C_A \tau/(2\pi R_\mathrm{m}) \rfloor = \lfloor  \tau/t_0 \rfloor$ is the maximum number of toroidal turns the Alfv\'en wave front has had time to make, with $t_0$ the time for the Alfv\'en wave to propagate one turn around the torus, accounting for the fact that emission is from both ends of the cloud. The notation $\lfloor x\rfloor$ gives the greatest integer less than or equal to $x$.

\subsubsection{Fraction of the cloudlet cross section connected to the opposite side}
\label{fraction}
We will now calculate the fraction $P_i$ of the cloudlet cross-section that connects to the opposite side during the $i^\mathrm{th}$ turn, assuming an irrational safety factor on the flux surface. We note though that rational flux surfaces have been shown to affect the ablation process -- owing to the smaller reservoir of hot electrons that can enter the pellet cloud -- and can also produce a drift braking effect, caused by the effective short-circuiting of the potential variation along field lines \citep{Commaux_2010,Sakamoto_2013}. Our model does not account for this, and as such it may provide a conservative upper estimate of the drift distance. As we shall see, most of the contribution to the current comes from terms with $i \gg 1$, i.e., from field lines that encircle the torus many times before connecting the two ends of the cloud. According to Weyl's lemma \citep{Helander_2014}, whether a given field line starting from one side of the cloud connects to the other side in a large number of turns is essentially random. We can thus speak of the probability of such a connection, and this probability depends on the fraction of the poloidal cross-section that the cloudlet covers, which is $\Delta y/(\pi r)$, where $r$ is the characteristic minor radius at the cloudlet position. Therefore,
    the total connected fraction $P_\mathrm{con}^\mathrm{tot}$ increases between turn $N$ and $N+1$ in the following way:
        \begin{equation}
            P_\mathrm{con}^\mathrm{tot}(N+1)-P_\mathrm{con}^\mathrm{tot}(N) = \frac{\Delta y}{\pi r}(1-P_\mathrm{con}^\mathrm{tot}(N)).
        \end{equation}
    The solution of this difference equation is
        \begin{equation}
            P_\mathrm{con}^\mathrm{tot}(N) = 1 - \left(1-\frac{\Delta y}{\pi r}\right)^N,
        \end{equation}
   and we can now express
        \begin{equation}
            P_i = P_\mathrm{con}^\mathrm{tot}(i+1)-P_\mathrm{con}^\mathrm{tot}(i) = \frac{\Delta y}{\pi r}\left(1-\frac{\Delta y}{\pi r}\right)^i.\label{eq:P_i}
        \end{equation}
This estimate is  consistent with figure 3 in \citep{Pegourie2006}. We can also express the fraction $P_A$ (determining the size of the Alfv\'en current) as $P_A = 1 - P_\mathrm{con}^\mathrm{tot}$.

Combining equation \eqref{iohmic}, \eqref{eq:iohmic_simple} and \eqref{eq:P_i}, the ohmic current contribution can now be expressed as
\begin{equation}
    I_{\|, \mathrm{ohm}} = \sum _{i=1}^N P_i\sigma _{\|} \frac{E_y\Delta y^2\Delta R}{2\pi R_\mathrm{m} i} = \frac{E_y\Delta R}{R_\mathrm{eff}},
\end{equation}
with the inverse effective resistivity $1/R_\mathrm{eff}$ given by
\begin{equation}
\frac{1}{R_\mathrm{eff}} = \sum _{i=1}^N P_i\sigma _{\|} \frac{\Delta y^2}{2\pi R_\mathrm{m} i} = \sigma _{\|}\frac{\Delta y^3}{2\pi R_\mathrm{m}\pi r }\sum _{i=1}^N \frac{1}{i}\left(1-\frac{\Delta y}{\pi r}\right)^i.
\label{eqReff}
\end{equation}
For $N\rightarrow \infty$ we may use $\sum _{i=1}^\infty (1-x)^i/i=-\ln{x}$, giving
\begin{equation}
    \frac{1}{R_\mathrm{eff}}=\sigma _{\|}\frac{\Delta y^3}{2\pi^2 R_\mathrm{m} r}  \ln{\frac{\pi r}{\Delta y}}.
    \label{Reffinf}
\end{equation}

Concerning when the $N\rightarrow \infty$ limit is meaningful to take, we must appreciate that depending on the resistivity of the cloud, the cloud may or may not be frozen into the magnetic field, which determines whether field lines are dragged along with the cloud, or the field lines slip with respect to the cloud\footnote{Fig.~3 of \citep{Hoare_2019} is a nice example from the scrape-off layer filament literature of exploring this transition numerically.}. It is worth re-iterating that the generation of Alfv\'{e}n waves by the propagating potential perturbation -- and so the existence of Alfv\'{e}n resistivity -- does not require the field lines to be frozen in on the drift time scale.    

As we will see later, the number of connected turns $N$ becomes large during the drift motion, so that taking $N\rightarrow \infty$ is a valid approximation for the majority of the drift motion, as long as the magnetic field diffusion is slow enough (i.e.~the cloud temperature is high enough) that the cloud does not become disconnected from the field lines where the electrostatic potential has been set up. 

The picture becomes more complicated if the magnetic field diffusion time scale is fast compared to the drift motion. This is typically the case for low cloud temperatures (e.g.~pellets doped with highly radiating impurities), where the conductivity in the cloud is low and the resistive diffusion coefficient is large. In this case, the potential along a given field line will not only be determined by the local cloud properties, but will be affected by all material which has drifted past the field line under consideration. When pellet material first arrives, the Alfvén current dominates. On the other hand, long after the ablation flow started to cross a given field line, the potential along this field line will reach a quasi-stationary profile similar to the case when the field line remains frozen into the cloud for a long time, and thus $N\rightarrow \infty$ also in this case. As the pellet motion is typically slow compared to the other processes of interest, the latter limit should dominate for the majority of the ablated material in most cases even for low cloud temperatures. 

The fraction of connected field lines, $P_A$, converges somewhat slower than the effective resistivity $R_\mathrm{eff}$. We therefore keep $N$ finite in the expression for $P_A$ for hot clouds. For cold clouds, for the first material drifting past a new part of the background plasma $N$ remains equal to zero, and $P_A=1$. However, as the potential reaches its quasi-stationary value, $P_A \rightarrow 0$ for the whole drift motion (i.e.~the parallel current will be dominated by the ohmic component).

 \subsection{Current balance}
We are now finally ready to sum up the various contributions to the current balance and obtain an equation for $E_y$ in terms of the parameters characterising the pellet cloud and the background plasma. From equation (\ref{integral}) we have
\begin{equation}
\begin{split}
    0 &= \frac{I_{\nabla B} + I_{\dot{\mathbf{E} }} + I_{\|, A} + I_{\|, \mathrm{ohm}}}{\Delta R} \\
    &= -\frac{4(\bar{n}T-L_\mathrm{cld}n_\mathrm{bg}T_\mathrm{bg})q}{BL_\mathrm{cld}}\sin{\left(\frac{L_\mathrm{cld}}{2qR_\mathrm{m}}\right)} + \frac{\bar{n}\langle m_i \rangle}{(1+\langle Z \rangle)B^2} \frac{dE_y}{dt} + 2P_A \frac{E_y}{R_A} + \frac{E_y}{R_\mathrm{eff}},
\end{split}
\label{eq:current_balance}
\end{equation}
Note that the factor $\sin{\left(\frac{L_\mathrm{cld}}{qR_\mathrm{m}}\right)}$ will start to oscillate when $t\sim q R_\mathrm{m}/c_s$, as $L_\mathrm{cld} \sim c_s t$, and the amplitude of the term in which this appears in \eqref{eq:current_balance} decreases as $1/L_\mathrm{cld}\propto 1/t$; this oscillation, together with the pressure equilibration (which occurs when $\bar{n}T=L_\mathrm{cld}n_\mathrm{bg}T_\mathrm{bg} $), effectively sets the time scale of the drift duration and eventually leads to a finite displacement for the drift. Also note that $c_\mathrm{s}t_0/qR_\mathrm{m}\sim c_\mathrm{s}/C_A$ is small for typical fusion plasma parameters, meaning that $N$ becomes large during the drift duration, motivating us to take the upper limit of the sum in (\ref{eqReff}) to be infinite, when calculating $R_\mathrm{eff}$.  

If the plasmoid and background plasma properties do not depend on $E_y$, equation \eqref{eq:current_balance} becomes a linear first-order ordinary differential equation in $E_y$, which can be written in the form
\begin{equation}
    \frac{dE_y}{dt} + g(t) E_y = f(t),
\end{equation}
with
\begin{equation}
    g(t) = \frac{(1+\langle Z \rangle)B^2}{\bar{n}\langle m_i \rangle}\left(2P_A \frac{1}{R_A} + \frac{1}{R_\mathrm{eff}}\right)
\end{equation}
and
\begin{equation}
    f(t) = \frac{4(1+\langle Z \rangle)B}{\langle m_i \rangle L_\mathrm{cld}}\left(T-\frac{L_\mathrm{cld}n_\mathrm{bg}}{\bar{n}}T_\mathrm{bg}\right)q\sin{\left(\frac{L_\mathrm{cld}}{2qR_\mathrm{m}}\right)}
\end{equation}
This equation can be solved by using an integrating factor $e^{G(t)}$, so
\begin{equation}
        E_y = e^{-G(t)}\left(E_{y0} + \int _0^t e^{G(t)}f(t)dt\right),
\end{equation}
where $E_{y0}=E_y(t = 0)$ and $G(t) = \int _0^t g(t)dt$. For a hot cloud, we have
\begin{equation}
\begin{split}
        G(t) &= \frac{(1+\langle Z \rangle)B^2}{\bar{n}\langle m_i \rangle}\left(2\left[1-\left( 1-\frac{\Delta y }{\pi r}\right)^{N+1}\right]\frac{\pi r}{\Delta y}  \frac{t_0}{R_A} + \frac{t}{R_\mathrm{eff}}\right) \\
    &= 2\left[1-\left( 1-\frac{\Delta y }{\pi r}\right)^{N+1}\right]\frac{\pi r}{\Delta y}  \frac{R_\mathrm{eff}}{R_A}\frac{t_0}{t_\mathrm{acc}} + \frac{t}{t_\mathrm{acc}},
\end{split}
\label{eq:G_full}
\end{equation}
where we have defined
\begin{equation}
    t_\mathrm{acc} = \frac{\bar{n}\langle m_i \rangle R_\mathrm{eff}}{(1+\langle Z \rangle)B^2}.
    \label{eq:t_acc}
\end{equation}
This is the characteristic acceleration time scale if the ohmic current dominates over the Alfv\'en current; if $R_\mathrm{eff}/R_A$ is small (corresponding to a hot background plasma), or in the case of a cold cloud long after the ablation flow started to cross the local field line, the expression (\ref{eq:G_full}) reduces to
\begin{equation}
     G(t) = \frac{t}{t_\mathrm{acc}}.
     \label{eq:G_simple}
\end{equation}
For a cold cloud shortly after the ablation flow started at the local field line, where $P_A = 1$, equation \eqref{eq:G_full} reduces to the same expression but with $R_\mathrm{eff}$ replaced with $R_A$ in the expression for $t_\mathrm{acc}$.

Finally, as the radially outward drift velocity of the cloudlet is due to the $\mathbf{E}\times \mathbf{B}$ motion, it can be estimated as $E_y /B$. Time integration leads to an expression for the net radial displacement 
\begin{equation}
    \Delta r = \frac{1}{B}\int _{0}^{\infty} E_y \mathrm{d}t.
    \label{timeintegralDeltar}
\end{equation}

\section{Parallel expansion and the final drift displacement}
\label{sec:parallelexpansion}
In this section, we complete the description of the pellet cloud by defining the density source resulting from pellet ablation. We then evaluate the drift of the pellet cloud, demonstrating its dependence on pellet composition and background plasma temperature.
\subsection{Model for the line-integrated density and cloud expansion}

The line-integrated density can be determined based on an estimate of how many particles the cloud contains when it detaches from the pellet source. The latter can be obtained as the product of the ablation rate and the time during which the pellet source is ablating inside the cloud.

A widely used  estimate for the mass ablation rate is given by
$$
\mathcal{G}=\lambda(X)\left(\frac{T_\mathrm{keV}}{2}\right)^{5/3}\left(\frac{r_p}{r_{p0}}\right)^{4/3}n_{e20}^{1/3},
$$
where $\lambda(X)=[27.1+\tan{(1.48X)}]/1000 \; \rm kg/s$, $X$ is the molecular fraction of deuterium in the pellet, $T_\mathrm{keV}$ is the background electron temperature in keV, $r_p$ is the pellet radius, $r_{p0}=2 \;\rm mm$ and $n_{e20}$ is the background electron density in units of $10^{20}\;\rm m^{-3}$ \citep{Parks_TSDW}. This expression is based on a version of the Neutral Gas Shielding (NGS) model~\citep{Parks_Thurnbull} that allows the pellet material to have both hydrogenic and noble gas components.

To determine the average detachment time (during which the pellet source contributes to the cloud), we estimate the initial acceleration $\dot{v}_0=\dot{v}(t=0)=\dot{E_y}(t=0)/B$  by balancing the first two terms in the current balance equation  \eqref{eq:current_balance}. The last two terms in \eqref{eq:current_balance} can be neglected, since in the initial phase, $E_y$ is small.
The time derivative of the electric field then becomes $$
\frac{dE_y}{dt}=\frac{2B(1+\langle Z \rangle)}{\bar{n}\langle m_i \rangle R_\mathrm{m}}\left(\bar{n} T-L_\mathrm{cld} n_\mathrm{bg}T_\mathrm{bg}\right),
$$
so that the initial acceleration is
\begin{equation}
    \dot{v}_0=\frac{1}{B}\frac{dE_y}{dt}=\frac{2(1+\langle Z \rangle)}{\langle m_i \rangle R_\mathrm{m}}\left(T_0 - \frac{n_\mathrm{bg}}{n_0}T_\mathrm{bg}\right),
\end{equation}
where $n_0=\bar{n}/L_\mathrm{cld}$ is the initial cloud density and $T_0$ is the initial temperature.

Initially, the pellet cloud is neutral, and it expands radially, but as soon as the particles are ionized, the expansion will continue along the magnetic field lines. The initial parallel expansion takes place at the speed of sound at a temperature of approximately $T_0$ (which is of the order of a few eV), and starts from a spherical cloud of cross section area $\pi \Delta y^2$. We can therefore estimate the density from mass conservation according to
\begin{equation}
    \mathcal{G} = 2n_0\langle m_i \rangle c_s(T_0)\pi \Delta y^2 \Rightarrow n_0 = \frac{\mathcal{G}}{2\langle m_i \rangle c_s(T_0)\pi \Delta y^2}.
\end{equation} 

The average distance the ablated material must drift before it exits the initial expansion tube around the pellet is $\Delta y$. Assuming that the initial motion has a constant acceleration we find
$$
\Delta y=v_0 t_\mathrm{det} + \dot{v}_0 t_\mathrm{det}^2/2,
$$
and
the average detachment time thus becomes
\begin{equation}
    t_\mathrm{det} = -\frac{v_0}{\dot{v}_0}+\sqrt{\left(\frac{v_0}{\dot{v}_0}\right)^2+\frac{2\Delta y}{\dot{v}_0}},
\end{equation}
where $v_0=v_p$ is the initial cloud velocity relative to the pellet, which is equal in magnitude but opposite in sign to the pellet velocity (assuming the cloud would be frozen in to the field lines where it was ablated in the absence of the $E\times B$ acceleration).
During the detachment time $t_\mathrm{det}$ the cloud expands to a length
\begin{equation}
    L_c = 2c_s(T_0)t_\mathrm{det},
\end{equation}
which serves as an initial condition for the cloud length for the remainder of the drift motion.
The line-integrated density is thus \begin{equation}
    \bar{n} = \frac{\mathcal{G}t_\mathrm{det}}{\langle m_i \rangle\pi \Delta y^2}.
\end{equation}

When the cloud detaches from the pellet, the temperature initially rises quickly due to the heating from hot electrons in the background plasma, but the details depend on the density and composition of the cloud. At low densities, the mean free path of the background-plasma electrons in the cloud is longer than the cloud itself. These electrons thus pass through the cloud and heat it relatively uniformly \citep{aleynikov_breizman_helander_turkin_2019,runov_aleynikov_arnold_breizman_helander_2021,Arnold_2021}. Most of the literature, however, considers the opposite limit of a dense cloud, where the stopping power is so great that the hot electrons cannot easily pass through it. We only consider this case but note that it becomes inapplicable at low cloud densities and high background-plasma temperatures. The heating also depends on the pellet composition; if the pellet contains even a small amount of a high-Z radiative component, the radiation from the pellet cloud quickly reaches a balance with the heating from the background plasma, and therefore the temperature rises far more slowly \citep{Matsuyama2022}. 
For pure hydrogen pellets, on the other hand, the radiation is too weak to have a major impact on the energy balance, and then the cloud temperature will relatively quickly increase to several tens of eV. 

The dependence of the cloud temperature on the background plasma temperature will be rather weak, as a higher background plasma temperature means both an increased heating and an increased ablation rate, giving more particles to absorb and, in the case of a high-Z-doped pellet, radiate away the energy. The heat flux scales as $q_\mathrm{bg}\sim T_\mathrm{bg}^{3/2}$ (neglecting any scaling of the cloud cross section area with the temperature), and the ablation rate scales as $\mathcal{G}\sim T_\mathrm{bg}^{5/3}$, so that the cloud temperature scales as $T\sim q_\mathrm{bg}/\mathcal{G}\sim T_\mathrm{bg}^{-1/6}$, i.e. a very weak scaling. Typical values for the cloud temperature, based on the results presented in ~\cite{Matsuyama2022}, are  $T=5\,\rm eV$ for neon doped pellets and $T=30\,\rm eV$ for pure hydrogen pellets. In the following we will assume the cloud temperature is constant during the drift motion and is independent of the background plasma temperature. This approximation is, of course, quite crude but not more so than other simplifications we have employed. 

Finally, as long as the cloud pressure is much higher than the background plasma pressure, the cloud will expand by approximately the speed of sound inside the cloud, $c_s\approx \sqrt{(\gamma _e\langle z \rangle + \gamma _i)T/\langle m_i \rangle}$, with $\gamma _e = 1$ and $\gamma _i = 3$, and will slow down when the cloud pressure becomes comparable to the background plasma pressure. Here we assume that the expansion speed is equal to $c_s$ as long as the cloud pressure is higher than the background plasma pressure, and then stops immediately when the cloud pressure becomes equal to the background plasma pressure, i.e.
\begin{equation}
    L_\mathrm{cld}\approx L_c + 2c_s\mathrm{min}(t,t_\mathrm{pe}),
\end{equation}
where the pressure equilibration time is 
\begin{equation}
    t_\mathrm{pe} = \frac{T\bar{n}}{2c_s n_\mathrm{bg}T_\mathrm{bg}}.
    \label{eq:t_pe}
\end{equation}
With the parallel dynamics model presented here, we have all the details needed to evaluate the electric field inside the cloud, and the drift displacement  can be calculated by evaluating the integral in (\ref{timeintegralDeltar}). Analytical expressions for the drift displacement in various limits are given in Appendix~\ref{appendix}.

Before evaluating the above expressions for the drift displacement for an ITER-like scenario, as a validation exercise, we evaluate the drift distance for parameters representative of a JET shattered pellet injection (SPI) experiment studied by \cite{Kong_EPS2022} using the JOREK code; discharge \#96874. Here, the drift displacement was accounted for by imposing a fixed prescribed shift between the ablating pellet shard and the location where the ablated material is eventually deposited.  A shift of $\Delta r = 30\,\rm cm$ was found to yield the best agreement to the experimental density evolution. We choose this case due to the availability of an estimate of the total drift displacement under experimental conditions as close as possible to the ITER-like scenario studied below in section \ref{sec:ITER_calc}.

The injected pellet consisted of $1.6\cdot 10^{23}$ deuterium atoms and was shattered into $\sim 300$ shards, which, using the Parks size distribution model \citep{ParksGA2016}, gives a characteristic shard radius of $r_p=0.6\,\rm mm$. However, as shards of larger volume contribute more to the density build-up -- that was matched to the experiment -- we consider a representative pellet shard radius of $r_p = 1\,\rm mm$ in our estimate. Lacking  values for $\Delta y$ and $T$ for the specific case, we set $\Delta y = 1.25$ cm and $T=30$ eV based on simulation results  by \cite{Matsuyama2022} of the same ITER-like scenario as the one studied below in section \ref{sec:ITER_calc}. Finally, using the representative geometrical and background plasma parameters $v_0 = 300\,\rm m/s$, $q=1.5$, $B = 3.45\,\rm T$, $R_\mathrm{m} = 3.5\,\rm m$, $r = 0.5\,\rm m$, $n_\mathrm{bg} = 8.5\cdot 10^{19}\,\rm m^{-3}$ and $T_\mathrm{bg}=7\,\rm  keV$ \citep{Kong_EPS2022}, we arrive at an estimated drift displacement of $\Delta r = 28\,\rm cm$, in good agreement with the value found to match the experimental data. Although this estimate may be altered by a factor $\sim 2$ within reasonable ranges of the relevant parameters, this result suggests that the present model is sufficiently accurate for order-of-magnitude estimates and qualitative studies, such as those performed for an ITER-like scenario in the next subsection.

\subsection{Calculation of the drift distance in an ITER-like scenario}
\label{sec:ITER_calc}
We now evaluate the above expressions for the drift displacement for parameters of interest in an ITER-like scenario, similar to that studied by \cite{Matsuyama2022}. In this scenario, the drifting pellet cloud is ablated from a pellet shard with radius $r_\mathrm{p} = 2\,\rm mm$ located at major radius $R_\mathrm{m} = 5\,\rm m$ and travelling with a speed of $v_0 = 500\,\rm m/s$ towards the high field side (i.e.~the injection is from the low-field side). We also assume that the cloud is initially stationary in the lab frame, so that $E_{y0} = 0$. The background plasma has a free electron density of $n_\mathrm{bg} = 10^{20}\,\rm m^{-3}$ and the magnetic field strength is $B = 5\,\rm T$. Moreover, we set $q = 1$, $\Delta y = 1.25\,\rm cm$, (based on simulation results by \cite{Matsuyama2022}) and the average charge for the neon is approximately $\langle Z_\mathrm{Ne} \rangle \approx 2$ at 5 eV. The background plasma temperature $T_\mathrm{bg}$ and the pellet composition will be varied.

\cite{Matsuyama2022} uses a model similar to that used by \cite{Pegourie2007}, adapted to mixed neon-deuterium pellets, including a Neutral Gas and Plasma Shielding (NGPS) model for the pellet ablation and a volume-averaged single-cell Lagrangian model for the parallel expansion. However, \cite{Matsuyama2022} only considers the early stages of the drift motion during the first $130\,\rm\mu s$ after the cloud has detached from the pellet, for a single isolated cloud, and does therefore not include the effect of ohmic currents and  rotational transform. Thus, the model by \cite{Matsuyama2022} accounts for the same physical mechanisms concerning the drift motion as ours in the case of a cold cloud shortly after the ablation flow has started to cross the local field lines\footnote{The effect of the rotational transform does not make a substantial difference during the first $130\,\rm\mu s$ in a large device such as ITER where $t_\mathrm{pol}$ typically ranges from a few hundred microseconds to a millisecond}. He concluded that the drift displacement is likely to be substantial compared to the plasma minor radius for pure hydrogen pellets, but will be strongly reduced in the presence of even a small amount of neon. Here, we attempt to reproduce this result in the corresponding limit, and then extend it by calculating the drift displacement after a long time, including the effect of ohmic currents.

\begin{figure}
    \centering
        \includegraphics[width = 0.49\textwidth]{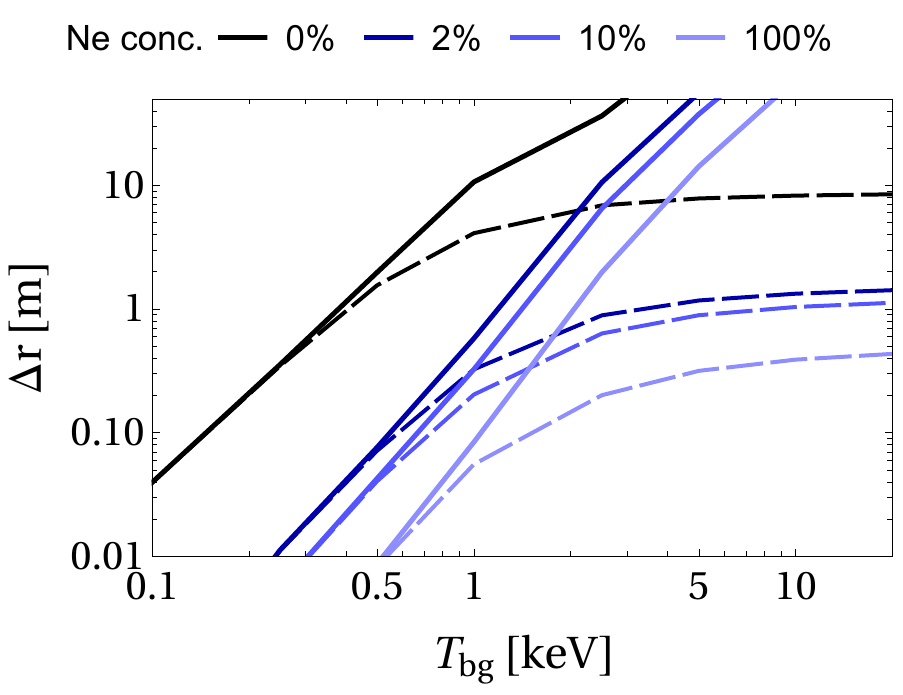}
    \put(-20,35){a)}
    \includegraphics[width = 0.49\textwidth]{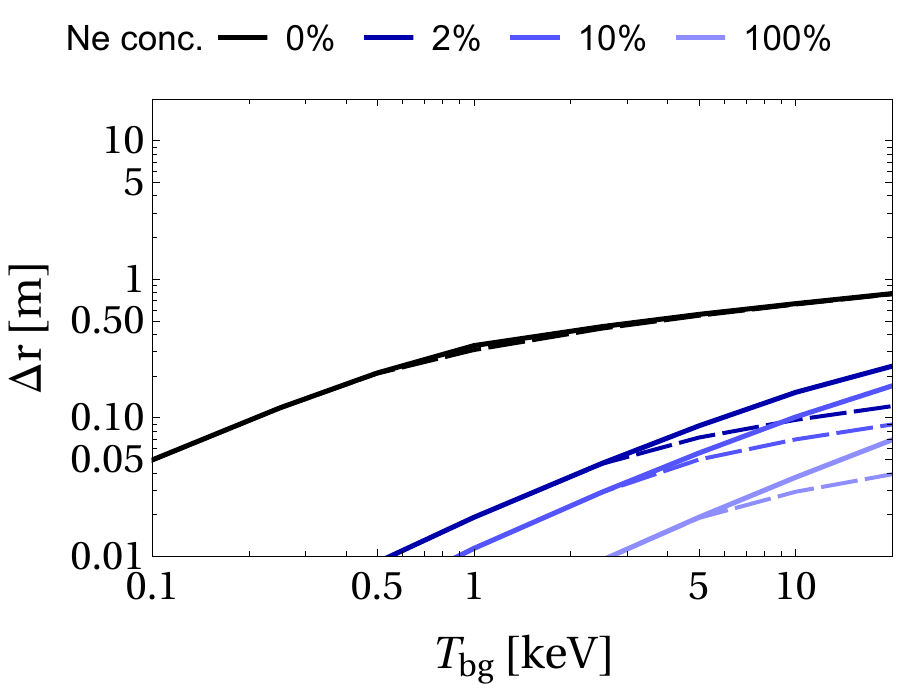}
    \put(-20,35){b)}
       \caption{Drift displacement as a function of background plasma temperature and pellet composition for cold clouds (30eV for pure hydrogen, 5eV otherwise), with different integration times and assumptions for the parallel current. In panel a) the parallel current is assumed to be purely Alfv\'enic (corresponding to when the ablation flow has just started to cross the local field lines), and in panel b) the parallel current  is assumed to be purely ohmic (corresponding to long after the ablation flow started to cross the local field lines). The solid lines correspond to performing the time integral of the drift velocity to $t=\infty$, as in (\ref{timeintegralDeltar}), the dashed lines are obtained by integrating only to $130\,\rm \mu s$.}
    \label{fig:drift_akinobu_comp}
\end{figure}

Figure \ref{fig:drift_akinobu_comp} shows the drift displacement for cold clouds (30eV for pure hydrogen, 5eV otherwise).
This is calculated by integrating \eqref{eq:E_anal} (leading to \eqref{eq:Dr_full_cold} if we integrate up to infinity), as a function of the background plasma temperature and pellet composition, with different integration times and assumptions regarding the ohmic currents. In panel a) we consider the case when the ablation flow has just started to cross the local field lines, i.e.~with the parallel current consisting only of the Alfv\'en current, and panel b) shows the results for long after the ablation flow started to cross the local field lines, i.e.~with the parallel current being purely ohmic. 

The dashed lines in panel a) are calculated with the assumption that the parallel current is purely Alfv\'enic, as was assumed by \cite{Matsuyama2022}, and the results are similar to those shown in figure 11 in \cite{Matsuyama2022} within an order unity factor, especially at high background plasma temperatures. The variation with both the background temperature and pellet composition agrees reasonably well. We see, however, that when we extend the integration time to infinity (solid lines), the drift displacement increases significantly at high background plasma temperatures, so that even clouds with 100\% neon would drift several meters in the absence of ohmic currents, although the drift displacement is not strongly affected for temperatures $\lesssim 1\,\rm keV$. This can be understood by considering that the pressure equilibration time becomes longer at high background plasma temperatures (see equation \eqref{eq:t_pe}), so that the cloud can drift a significant distance after the first $130\,\rm\mu s$. Moreover, in the absence of ohmic currents, the acceleration time scale is typically longer than $130\,\rm\mu s$, so that the cloud continues to gain speed even after this time frame. For low background plasma temperatures, on the other hand, the pressure equilibration time becomes shorter than $130\,\rm\mu s$ so the cloud does not drift significantly after this time. 

In panel b), where the parallel current is purely ohmic, we see that the drift displacement is reduced by about one order of magnitude when integrating up to $130\,\rm\mu s$ (compare with panel a), and about two orders of magnitude when integrating to infinity. The scaling with the background plasma temperature is also weaker, as anticipated above, because the resistivity determining the parallel current now scales with the background plasma temperature as $R_\mathrm{eff}\sim T_\mathrm{bg}^{-3/2}$, which mostly cancels the temperature scaling of the ablation rate $\mathcal{G}\sim T_\mathrm{bg}^{5/3}$ (there is some dependence on the background temperature left at lower background temperatures where the ratio of the cloud pressure and the background pressure is lower). Moreover, the effect of increasing the integration time beyond $130\,\rm\mu s$ is now much smaller than in the absence of ohmic currents. This follows as the acceleration time scale $t_\mathrm{acc}$ is much shorter, so that the cloud decelerates rather than accelerates after the first $130\,\rm\mu s$.

For neon-doped pellets, the drift displacement now ranges from a few cm up to $\sim 20$ cm at the highest relevant temperatures, which is small compared to both the plasma minor radius and the plume of shards in case of a SPI in an ITER-like scenario. The pure deuterium pellets, on the other hand, still have a drift displacement of tens of cm, which is a sizeable fraction of the plasma minor radius and comparable to the radial extent of the shard plume in case of an SPI. This result corroborates the conclusion made by \cite{Matsuyama2022}.

We now compare the results for the same plasma scenario as above using the expressions obtained with different limits and model assumptions. 
As we have seen in section~\ref{fraction}, for hot clouds (e.g.~pure deuterium pellets), the $N\rightarrow\infty$ limit of $R_\mathrm{eff}$ can be used while we keep $N$ finite in the expression of $P_A$. For cold clouds (e.g.~neon-doped pellets), in the long-time limit (as the potential reaches its quasi-stationary value), the Alfvén part of the current can be neglected ($P_A=0$).

\begin{figure}
    \centering
    \includegraphics[width = 0.49\textwidth]{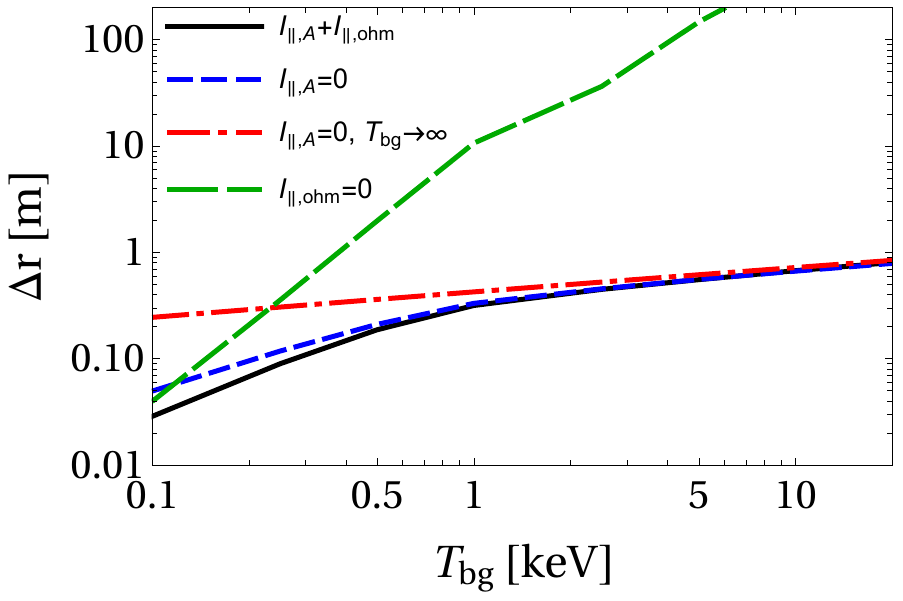}
    \put(-150,35){a)}
    \includegraphics[width = 0.49\textwidth]{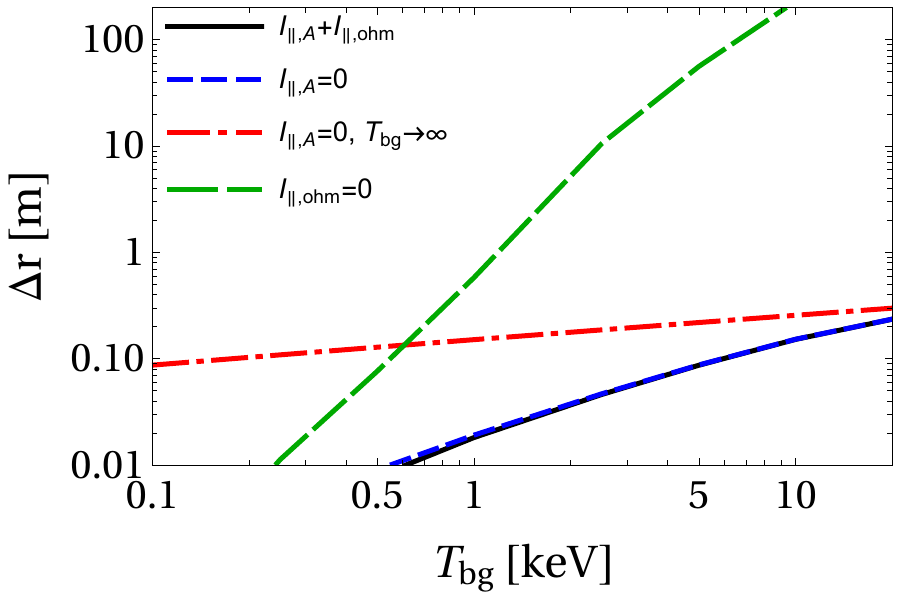}
    \put(-150,35){b)}
    \caption{Comparison of the drift displacement obtained with different limits and model assumptions, for a pellet consisting of a) $100\%$ deuterium and b) a mixture with $98\%$ neon and $2\%$ deuterium. Solid black: $I_\mathrm{\|,A}+I_\mathrm{\|,ohm}$, numerical integration of \eqref{eq:E_full}. Dashed blue: $I_\mathrm{\|,A}=0$, using \eqref{eq:Dr_full_cold}. Dash-dotted red: $I_\mathrm{\|,A}=0$ and taking $T_{\rm bg}\rightarrow \infty$ asymptotic behaviour, using \eqref{eq:Dr_simple}. Long dashed green:  $I_\mathrm{\|,ohm}=0$, using \eqref{eq:Dr_full_cold}, but with $R_{\rm eff}$ replaced by $R_{\rm A}$
    }
    \label{fig:model_comp}
\end{figure}

In figure \ref{fig:model_comp}, the full solution,
which contains both the $I_\mathrm{\|,A}$ and $I_\mathrm{\|,ohm}$ contributions obtained by numerically integrating \eqref{eq:E_full}, is shown by a black curve for a pure deuterium pellet (panel a) and a $2\%$ neon-doped one (panel b). We also consider the cases representing the long and short-time limits, in terms of the time passed after the ablation flow first started to cross the local field lines. In the short-time limit (green long-dashed curve) $I_\mathrm{\|,ohm}$ is neglected, and it is calculated by replacing $R_\mathrm{eff}$ by $R_A$ in equation \eqref{eq:Dr_full_cold}. The long-time limit (blue dashed curve) physically means that $I_\mathrm{\|,A}$ is neglected, and it is calculated using \eqref{eq:Dr_full_cold}. (Note that in the case of a cold cloud with a fast magnetic field diffusion time scale compared to the drift motion, in the long-time limit, the $I_\mathrm{\|,A}=0$ limit is expected to be accurate, as discussed at the end of Sec.~\ref{fraction}.) In addition, we also show results calculated using the simplified expression \eqref{eq:Dr_simple} (red dash-dotted), that represents the high-background-temperature asymptotic behaviour of the long-time limit.

We see that for both the pure deuterium and the neon-doped pellet, \ref{fig:model_comp}a and b, the long-time limit gives similar drift displacement to the general expression (compare dashed and solid), especially at high background-plasma temperatures. There is a discrepancy of $\lesssim 50\%$ at background temperatures of $T_{\rm bg}\sim 100\,\rm eV$ where the ohmic conductivity is rather low, but at these temperatures the displacement, and therefore the discrepancy, remains moderate. The overall good agreement reflects that the number of connections $N$ continuously increases with time in a hot cloud, so that the Alfv\'en conductivity is replaced by ohmic conductivity over a short period of time compared to the total drift time.

In the case of a pure deuterium pellet, \ref{fig:model_comp}a, we see that the high-background-temperature asymptotic form of the long-time limit (dash-dotted) approaches the more accurate expression \eqref{eq:Dr_full_cold} at $T_{\rm bg}\gtrsim 1\,\rm keV$, but the approach is much slower
in the doped-pellet case, \ref{fig:model_comp}b. This difference is due to the higher cloud temperature for a pure deuterium cloud, leading to a longer pressure equilibration time $t_\mathrm{pe}$ while the acceleration time $t_\mathrm{acc}$ remains only weakly affected by the background temperature, making the approximation $t_\mathrm{acc}/t_\mathrm{pe}\approx 0$ accurate at lower temperatures.

Finally, we find that the short-time limit (long-dashed curves in \ref{fig:model_comp}) 
typically gives unphysically large drift displacements, unlike the general expression and the long-time limit. Only at $T_\mathrm{bg}\lesssim 100\,\rm eV$ does the short-time-limit expression become comparable to or smaller than the long-time limit; then the ohmic conductivity of the background plasma becomes so low that the Alfv\'en conductivity starts to dominate. We note that at sufficiently low values of $T_\mathrm{bg}$, the short-time limit result starts to asymptotically approach the general expression (black curve), but that happens at very small, inconsequential, values of the drift displacement $\Delta r$.

\section{Discussion and Conclusion}
We have derived a semi-analytical model for the cross-field drift of an ionised cloud following a pellet injection in a tokamak. The model gives the radial drift velocity in terms of the background plasma and cloud properties, assuming the latter to be constant along the field lines inside the cloud. The main phenomena included in the model are the $\nabla B$ current causing the charge separation inside the cloud and the resulting $E\times B$ drift, the rotational transform, pressure equilibration, and the currents limiting the charge separation; the latter including the polarisation current and the currents exiting through the ends of the cloud parallel to the field lines, consisting of an Alfv\'enic and an ohmic contribution. In particular, we have developed a statistical model for the length of the field lines connecting the two ends of the cloud, and the corresponding effective resistivity for the Ohmic current flowing along those field lines.

We then derive semi-analytical expressions for the final drift displacement, combining our model for the cross-field drift with a simple analytical model for the cloud properties. We evaluate the resulting expressions in an ITER-like scenario similar to those studied by \cite{Matsuyama2022}, including a wide range of background plasma temperatures and different neon-deuterium mixtures for the pellet composition. Our results
are in reasonable agreement with those obtained by \cite{Matsuyama2022} in the corresponding limit, integrating only up to $130\,\rm\mu s$ after the cloud is detached from the pellet source and neglecting the ohmic part of the parallel current (corresponding to a cold cloud shortly after the pellet material has started to flow across a given field line). We then investigate the effect of adding the ohmic part of the parallel current and integrating to longer times. Without ohmic currents, the final drift displacement becomes unreasonably long, up to several tens (or even hundreds) of meters, while adding the ohmic current reduces the drift displacement by typically 1-2 orders of magnitude.

Our results suggest that a pure deuterium pellet injection in an ITER-like scenario is likely to be significantly affected by the radial drift displacement, and that a substantial part of the injected material may be expelled from the plasma. On the other hand, a neon-doped pellet injection will likely be significantly less affected by the drift displacement. This result corroborates the conclusion made by \cite{Matsuyama2022}.

Note, however, that even a relatively small drift displacement can have a significant effect on the ablation and density profile \cite{Vallhagen_MSc}. The reason is that even a small drift means that the pellet will not feel its own cooling effect on the background plasma, which otherwise provides a self-regulating feedback mechanism which decreases the ablation rate. Even a small drift therefore makes the pellet, or pellet shards, ablate faster, so that they deposit more of their material earlier along their trajectories. This applies especially to injections from the low field side, as in that case the drift will displace the ablated material behind the ablating source. On the other hand, an injection from the high field side will displace the ablated material in front of the pellet or pellet shard, so that it feels the effect of its own cooling along its trajectory. The importance of this effect also depends on the magnetic-field strength, which regulates the transverse dimension of the pellet cloud, and on the injection velocity of the pellet. The effect in question is most important if the field is strong and the pellet velocity is small. 

In the case of an SPI in an ITER-like scenario, the plume of shards typically extends over several decimetres. Thus, in the case of a neon-doped pellet, our results indicate that the shards will still feel the cooling of the background plasma from most shards ahead of them, even for an injection from the low field side. 
For a pure deuterium SPI, on the other hand, the drift displacement will likely be longer than the extent of the plume of shards, which might increase the ablation significantly, especially for an injection from the low field side. A quantitative assessment of the effect of the drift displacements calculated by the model presented here would require coupling
to a model for the full injection dynamics and response of the background plasma, which is outside the scope of the present work.

The accuracy of the results presented in this paper is also limited by a number of simplifications, primarily in the model for the parallel expansion and cloud properties. In particular, the cloud properties are assumed to be constant along the field lines inside the cloud, and the energy balance and temperature evolution is modelled using only a constant, representative value for the cloud temperature. While the cloud temperature remains rather low and constant for a neon-doped pellet due to the high radiated power, the temperature will vary significantly during the drift motion for a pure deuterium pellet; indeed, the discrepancy compared to the results obtained by \cite{Matsuyama2022} is larger for a pure deuterium pellet. The quantitative accuracy of the present model could therefore be significantly improved by combining the present model for the cross-field drift with a more advanced model for the cloud properties, which is outside the scope of the present work. 
\appendix
\section{Expression for the drift displacement in relevant limits}
\label{appendix}
 It is convenient to introduce the expansion time scale $t_\mathrm{exp} = L_c/(2c_s)$ and the time $t_\mathrm{pol}$ it takes the cloud to expand a poloidal angle of one radian. 
We also introduce the normalised time variable $t' = t/t_\mathrm{acc}$ and normalise the other time scales accordingly, also denoted with a prime, and introduce the shifted normalised time variable $t'' = t' + t_\mathrm{exp}/t_\mathrm{acc}$. In terms of these variables, the electric field inside the cloud can be expressed as
\begin{equation}
\begin{split}
    E_y &= E_{y0}e^{-G(t')} + \frac{2(1+\langle Z \rangle)BTq}{\langle m_i \rangle c_s}\times\\
    &e^{-G(t')}\int _0^{\mathrm{min}(t', t'_\mathrm{pe})}e^{G(\tilde{t}')}\left(\frac{1}{\tilde{t}''}-\frac{1}{t'_\mathrm{pe}}\right)\sin{\left(\frac{\tilde{t}''}{t'_\mathrm{pol}}\right)}d\tilde{t}'\\
    &=E_{y0}e^{-G(t')} + \frac{2(1+\langle Z \rangle)BTq}{\langle m_i \rangle c_s}\mathcal{E}\left(t'_\mathrm{pe}, t'_\mathrm{exp}, t'_\mathrm{pol}, \frac{R_\mathrm{eff}}{R_A}, t'\right),
\end{split}
\label{eq:E_full}
\end{equation}
where $\tilde{t}'' = \tilde{t}' + t_\mathrm{exp}/t_\mathrm{acc}$ and $\mathcal{E}$ is a dimensionless function of the time variable $t'$ with four dimensionless parameters. However, not all four parameters are relevant in all cases. If, for instance the ohmic currents dominate over the Alfv\'en current (such as for a hot background plasma or for a cold cloud long after the ablation flow started to cross the local field line), we can set $R_\mathrm{eff}/R_A = 0$. In this case, $\mathcal{E}$ can be expressed in closed form as
\begin{equation}
\begin{split}
    &\mathcal{E}\left(t'_\mathrm{pe}, t'_\mathrm{exp}, t'_\mathrm{pol}, 0, t'\right)\\
    &= e^{-t'}\left\{ e^{-t_\mathrm{exp}}\mathfrak{Ei}\left[\left(1+\frac{i}{t'_\mathrm{pol}}\right)t''\right]- \frac{1}{t'_\mathrm{pe}} e^{t'}\frac{\sin{\left(\frac{t''}{t'_\mathrm{pol}}\right)} - \frac{1}{t'_\mathrm{pol}}\cos{\left(\frac{t''}{t'_\mathrm{pol}}\right)}}{1+{t'}_\mathrm{pol}^{-2}}\right\}_0^{\mathrm{min}(t', t'_\mathrm{pe})}\\
    &= e^{-t'}\left(\epsilon\left(t'_\mathrm{pe}, t'_\mathrm{exp}, t'_\mathrm{pol}, \mathrm{min}(t', t'_\mathrm{pe})\right) - \epsilon\left(t'_\mathrm{pe}, t'_\mathrm{exp}, t'_\mathrm{pol}, 0)\right)\right),
\end{split}
\label{eq:E_anal}
\end{equation}
with
\begin{equation*}
    \mathfrak{Ei}[x] = \frac{1}{2i}\left[\mathrm{Ei}(x) - \mathrm{Ei}(x^*)\right],
\end{equation*}
where $\mathrm{Ei}$ is the exponential integral function, $i$ is the imaginary unit, an asterisk superscript denotes complex conjugate, and we defined the expression within the curly bracket in equation \eqref{eq:E_anal} as $\epsilon$. Integrating equation \eqref{eq:E_anal}, we get the following expression for the drift displacement:
\begin{equation}
\begin{split}
    \Delta r &= \frac{E_{y0}}{B}t_\mathrm{acc} + \frac{2(1+\langle Z \rangle)Tq}{\langle m_i \rangle c_s}t_\mathrm{acc}\int _0^\infty \mathcal{E}\left(t'_\mathrm{pe}, t'_\mathrm{exp}, t'_\mathrm{pol}, 0, t'\right)dt'\\
    &= v_0t_\mathrm{acc} + \frac{4\bar{n}TR_\mathrm{eff}q}{B^2c_s}\left\{\epsilon\left(t'_\mathrm{pe}, t'_\mathrm{exp}, t'_\mathrm{pol}, t'_\mathrm{pe}\right) e^{-t'_\mathrm{pe}}-\epsilon\left(t'_\mathrm{pe}, t'_\mathrm{exp}, t'_\mathrm{pol}, 0\right)\right.\\
    &\left.+e^{-t'_\mathrm{exp}}\left[e^{-t'}\left\{e^{t''} \mathfrak{Ei}\left[i\frac{t''}{t'_\mathrm{pol}}\right]-\mathfrak{Ei}\left[\left(1+i\frac{1}{t'_\mathrm{pol}}\right)t''\right]\right\}\right]_0^{t'_\mathrm{pe}}\right.\\
    &\left.+ \frac{1}{t'_\mathrm{pe}}\frac{1}{1+{t'}_\mathrm{pol}^{-2}} \left[t'_\mathrm{pol}\cos{\left(\frac{t''}{t'_\mathrm{pol}}\right)} + \sin{\left(\frac{t''}{t'_\mathrm{pol}}\right)}\right]_0^{t'_\mathrm{pe}}\right\},
\end{split}
\label{eq:Dr_full_cold}
\end{equation}
where $v_0 = E_{y0}/B$ is the speed of the pellet. In some relevant cases, $\mathcal{E}$ can be simplified further; for high background temperatures, $t_\mathrm{acc}/t_\mathrm{pe}\approx 0$. Moreover, the cloud length typically becomes much longer than the initial length $L_c$ in a very short amount of time, so that we can approximate $L_c/(c_st_\mathrm{acc})\approx 0$. In that case, $\mathcal{E}$ only depends on a single parameter $t_\mathrm{pol}/t_\mathrm{acc}$, and can be expressed as
\begin{equation}
\begin{split}
    \mathcal{E}\left(\infty, 0, t'_\mathrm{pol}, 0, t'\right)&= e^{-t'}\left\{\mathfrak{Ei}\left[\left(1+i\frac{1}{t'_\mathrm{pol}}\right)t'\right]\right\}_0^{t'}\\ &=e^{-t'}\left\{\mathfrak{Ei}\left[\left(1+i\frac{1}{t'_\mathrm{pol}}\right)t'\right] - \tan^{-1}{\frac{1}{t'_\mathrm{pol}}}\right\}.
\end{split}
\label{eq:E_anal_approx}
\end{equation}
The corresponding expression for the drift displacement becomes
\begin{equation}
    \begin{split}
        \Delta r &= \frac{E_{y0}}{B}t_\mathrm{acc} + \frac{2(1+\langle Z \rangle)Tq}{\langle m_i \rangle c_s}t_\mathrm{acc}\int _0^\infty \mathcal{E}\left(\infty, 0, t'_\mathrm{pol}, 0, t'\right)dt'\\
        &=v_0t_\mathrm{acc} + \frac{\pi\bar{n}TR_\mathrm{eff}q}{B^2 c_s}.
    \end{split}
    \label{eq:Dr_simple}
\end{equation}
Equations \eqref{eq:E_anal}-\eqref{eq:Dr_simple} apply also to a cold cloud shortly after the ablation flow has started to cross the local field line, but with $R_\mathrm{eff}$ replaced with $R_A$, in accordance with the corresponding change in the expression for $t_\mathrm{acc}$, equation \eqref{eq:t_acc}.

Note that an increased acceleration time-scale leads to a longer drift displacement, which might seem surprising as that means that it takes longer for the cloud to get up to speed. This is however compensated by the increased inertia, preventing the cloud from slowing down when the acceleration changes sign due to the sign change of the net $\nabla B$ current, when the sine factor in equation \eqref{eq:current_balance} becomes negative.

\section*{Acknowledgements} 
The authors are grateful to  E Nardon and A Matsuyama for fruitful discussions. This work was supported by the Swedish Research Council (Dnr.~2018-03911) and part-funded by the EPSRC Energy Programme [grant number EP/W006839/1]. The work has been carried out within the framework of the EUROfusion Consortium, funded by the European Union via the Euratom Research and Training Programme (Grant Agreement No 101052200 — EUROfusion). Views and opinions expressed are however those of the author(s) only and do not necessarily reflect those of the European Union or the European Commission. Neither the European Union nor the European Commission can be held responsible for them.
\bibliographystyle{jpp}
\bibliography{bibliography}

\end{document}